\documentclass[preprint,showpacs,preprintnumbers]{revtex4}

\usepackage{amsfonts}
\usepackage{amsmath}
\usepackage{amssymb}
\usepackage{graphicx}

\begin{document}
\title{Universal geometric classification of armchair graphene nanoribbons by their properties in a staggered sublattice potential}
\author{T. E. O'Brien}
\affiliation{School of Engineering Physics, University of
Wollongong, New South Wales 2552, Australia}
\author{C. Zhang}
\affiliation{School of Engineering Physics, University of
Wollongong, New South Wales 2552, Australia}
\author{A.R. Wright}
\affiliation{Institute for Theoretical Physics, University of Leipzig, D-04103 Leipzig, Germany}

\begin{abstract}
We demonstrate the topological properties of the band-gap of armchair graphene nanoribbons in a spatially varying staggered sublattice potential. Several general scaling laws are presented to quantify the band gap variation. It is found that all armchair nanoribbons are described by one of three distinct classes depending on their width, one of which is the well known massless Dirac condition, and the other two we call potentially gapless, and gapless-superlattice. We construct an effective theory which faithfully reproduces these results, and makes explicit the nature of the competing masses and overlap integrals across a particular sample. Finally we propose several systems on which these results should shed considerable light, and which have all already been experimentally realized.
\end{abstract}

\maketitle

The search for states with topological origins is not a new one. The domain wall states in polyacetylene \cite{poly} and their field theoretic counterparts\cite{jakiw}, as well as the TKNN invariant and the quantum Hall effect \cite{tknn} followed by the fractional quantum Hall effect and Chern-Simons theories \cite{zhk}, are condensed matter examples which each span several decades. Undoubtedly this ubiquitous field has recently enjoyed a renaissance due to the excitement surrounding topological insulators: time reversal invariant bulk insulators with a single gapless surface Dirac cone \cite{zhangrev}.

The community's interest having been piqued, it is not surprising that graphene, the canonical example material with a 2D massless Dirac cone, not to mention the first material to have been considered as a topological insulator \cite{km}, has been cut, bent, layered, crumpled, adsorbed, twisted and stretched in order to demonstrate new and exotic topologically non-trivial states \cite{soliton, loss, qahe, blg1,blg2,toprev}. This has been a remarkably successful task, and graphene continues to produce new and unexpected physics each time someone pokes or prods it in a new way.

The purpose of the current work is to do this once again. It is well known that quasiparticles in armchair graphene nanoribbons are, in general, massive. The two $K$ points in bulk two dimensional graphene are projected onto each other in momentum space when the system is reduced to a quasi-one dimensional one, allowing the two massless cones to mix, thus producing a gap \cite{timotony}. In the seminal paper by Brey and Fertig, this is easily seen by noting that the boundary conditions of the two sublattices must match in this special geometry \cite{bf}. To contrast, in the case of zig-zag ribbons only one sublattice constitutes each edge, thus preserving the gapless K point for all zig-zag edged ribbons. For this single crucial reason, zig-zag edged ribbons, samples, and interfaces are the usual candidate systems in which topologically borne states are proposed to exist \cite{semenoff}.

Here, we show that armchair edged ribbons with a staggered sublattice potential have a tunable mass-gap which, in the case of gapped ribbons can be decreased, or even closed, by varying the potential strength and gradient. In complete analogy with the occurrence of massless Dirac ribbons, we show that with increasing width, each effect is observed in every third ribbon. By constructing an effective theory which reproduces all tight-binding results, we show that this tunability arises from the separate, competing masses originating from the staggered sublattice potential and the quasi-one dimensional confinement. This effective theory further betrays an analogy with the quantum anomalous Hall effect in pseudo-spin space. We propose three realistic systems where these results may be observable: twisted bilayer graphene, chemically adsorbed single layer graphene, and lattice or substrate mismatched graphene.

The Hamiltonian which is the focus of the current work is $H = H_0 + V$, where $H_0$ is the tight-binding Hamiltonian of graphene which is given by

\begin{equation}
H_0 = \sum_i^{2W} \sum_j^3 t_{ij}c^\dag_{i+\hat{\delta_j}}c_i + \mathrm{c.c.}
\end{equation}

Where $t_{ij}\approx 3eV$ is the overlap integral of first nearest neighbouring sites $\langle i,j\rangle$, and is zero otherwise, and $j$ denotes the three nearest neighbour vectors. In this formalism, the width of a ribbon or superlattice unit cell is $W\sqrt{3}|\hat{\delta}|/2$. To create a ribbon we simply construct the appropriate unit cell and set $t=0$ when a first nearest neighbouring site leaves the edge of the ribbon, and to create a superlattice of ribbons with alternating sublattice potential, we adopt periodic boundary conditions instead. We also introduce a staggered sublattice potential across the width of the ribbon given by

\begin{equation}
V = \sum_i^{W/2} \mathrm{sgn}(i-i_W)m \bigl(c_{i}^{\dag} c_{i} - c_{i+\hat{\delta_1}}^\dag c_{i+\hat{\delta_1}}\bigr)
\label{pot}
\end{equation}

\begin{figure}[tbp]
\centering\includegraphics[width=8cm]{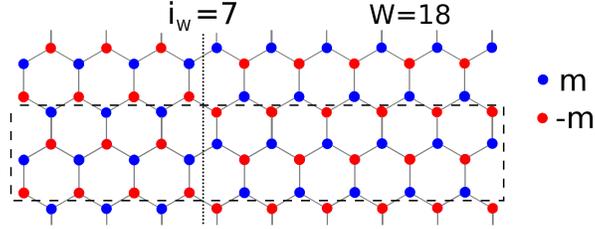}
\caption{A typical lattice structure (W/2=18, $i_W$=7). At the domain wall, the position of the alternating positive (red) and negative (blue) sublattice potentials is changed from A+/B- to A-/B+, where A and B denote the index of individual pairs. The AB pairs are chosen so as to be positioned along the axis of the ribbon. The unit cell is shown by the dashed line.}
\label{figsys}
\end{figure}

Where $m$ is the magnitude of the potential and $i_W$ marks the site in the unit cell where the potential changes sign, amounting to the location of  a domain wall. Although this sharp change in sign is a specific choice, we stress that none of our results are qualitatively changed by a smooth change in sign of the potential. In Fig. \ref{figsys} we show a typical ribbon that we are considering with the domain wall placed at a typical site $i_W$.

Upon diagonalising these systems, we first ask the following question: is it possible to completely collapse the band gap of an armchair ribbon by applying a periodic potential? The answer to this question is yes, but only in one \emph{sixth} of all possible ribbons, namely those where $(W-4)/6\in\mathbb{Z}$, and only when the $i_W = W/2$. We stress here that these ribbons are not Dirac ribbons ($(W-2)/3\in\mathbb{Z}$). Ribbons where $(W-1)/6\in\mathbb{Z}$ \emph{almost} possess this quality, however as they have an odd number of pairs in their unit cells, they cannot \emph{strictly} fulfill the requirements of equally sized domains ($W$ is odd, so $i_W = W/2\pm 0.5$), even though the effect is identical. Therefore these ribbons display extremely small minima of $\epsilon_{BG}$ compared to their non-Dirac counterparts ($W/3\in\mathbb{Z}$). We label this class of armchair ribbons potentially gapless (PG), as a tuned staggered sublattice potential has the ability to tune the band-gap to zero, or nearly zero.

\begin{figure}[tbp]
\centering\includegraphics[width=8cm]{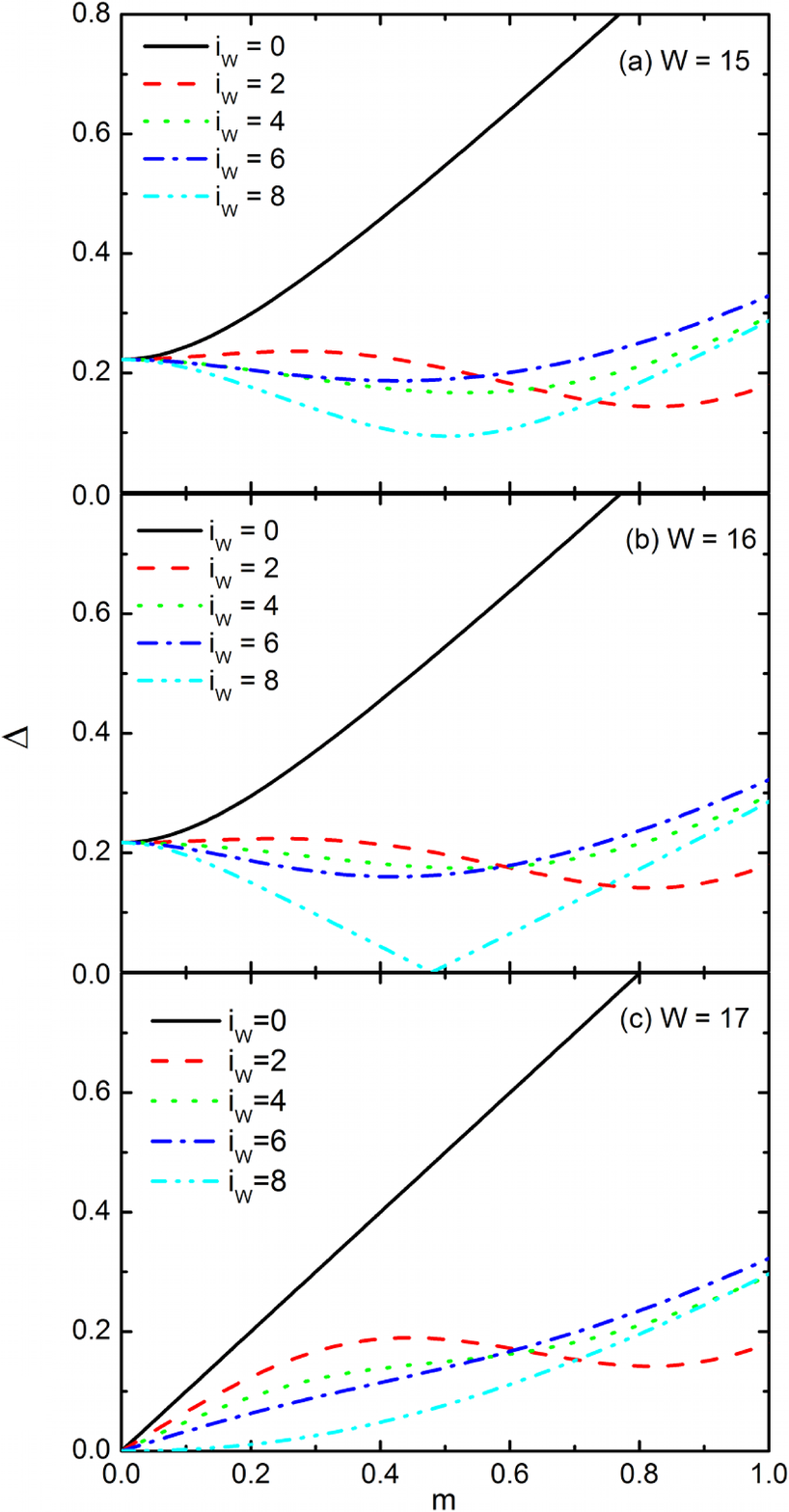}
\caption{Graphs of band gap trends with increasing m for (top to bottom) GS, PG (with even number of A-B pairs), and Dirac ribbons. Note the large dip with all ribbons when $i_{W}=2$, which is the smallest possible Dirac domain}
\label{massdependence}
\end{figure}

We now move on to the armchair superlattice. The system is now two-dimensional, and the lattice translation vectors are orthogonal. In general, the transverse component of the Hamiltonian is easily constructed as a two-site unit cell with lattice vectors trivially determined by $i_W$. The resulting bandstructure behaviour along the transverse direction is $\cos$-like. Due to the oscillating gap along the transverse direction, the band-gap position deviates from the $\Gamma$ point. In particular, the direct band gap position moves between $k_x = 0$, $k_x = 2\pi/3$ and $k_x = \pi$, where $\pi$ is the dimensionless edge of the Brillouin zone.

\begin{figure}[tbp]
\centering\includegraphics[width=8cm]{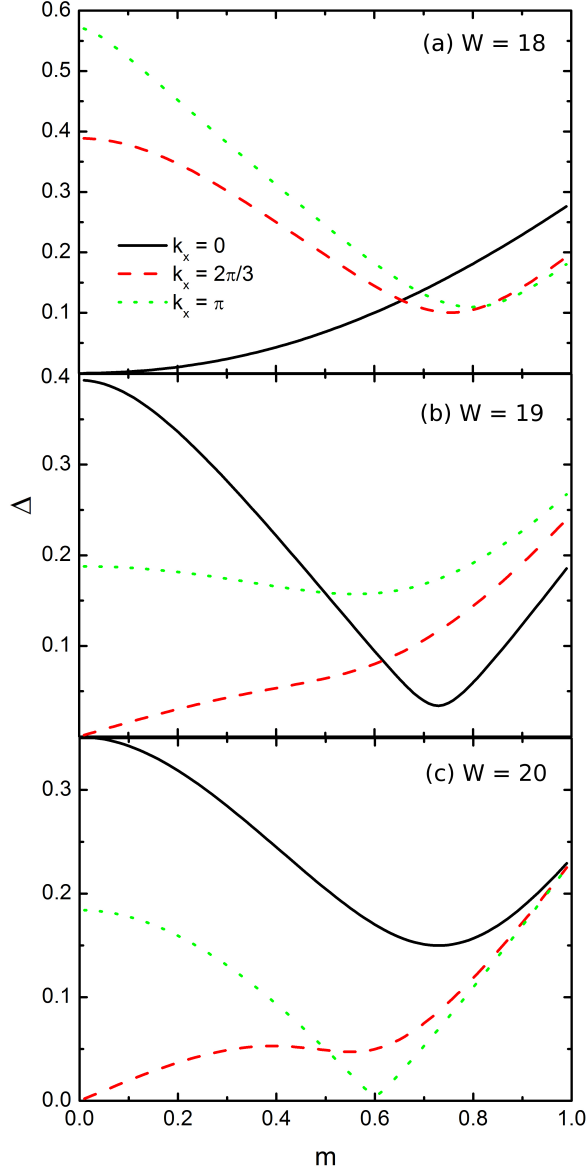}
\caption{The mass dependence of the band gap at the $\Gamma$ point, $k_x = 2\pi/3$, and $k_x = \pi$. The behaviour at the $\Gamma$ establishes the third universal classification of gapless superlattice `ribbons', which are the only armchair superlattices to become gapless as $m\rightarrow 0$. The other two points define the high-symmetry gapless points and band-inversion points, which are lifted for sufficiently large $m$. For each system, $i_W = 8$.}
\label{figsl}
\end{figure}

At the gamma point, Dirac and PG superlattices are gapped at $m=0$, however PG superlattices can close the gap at $m = \Delta$ as in the ribbon case. However the third, so-far uncategorised ribbon type here differs. This brings us to a definition for the third class of ribbons: those for which $W\in\mathbb{Z}$, are \emph{always} gapless at the $\Gamma$ point when $m=0$. Thus these superlattices are likened to armchair Dirac ribbons, in this sense. This is a particularly important class of ribbon, as they show that a very weak modulation across a 2D sample can close the gap at the gamma point, thus destroying the valley degeneracy of a 2D sheet. (We should point out that this is strictly the low $m$ limit, rather than $m = 0$, as the latter simply corresponds to an infinite graphene sheet with zone-folded (incorrect) energy bands). This third class of `ribbon' we call gapless superlattice (GS) ribbons. The gap at the gamma point as a function of $m$ is shown in Fig. \ref{figsl} by the solid black line. We have thus established that \emph{all} graphene ribbons fall into one of the three classes of Dirac ($(W-2)/3\in\mathbb{Z}$), potentially gapless ($(W-1)/3\in\mathbb{Z}$), and gapless superlattice ($W/3\in\mathbb{Z}$).

The second important point in the superlattice Brillouin zone is $2\pi/3$, whose behavior with increasing $m$ is shown in Fig. \ref{figsl} by the red dashed line. At this point, there is a band crossing between the top-most valence band, and the lowest conduction band, leading to two Dirac points at $m= 0$ for Dirac and PG superlattices, directly corresponding to the two $K$ points of a 2D sheet. As the mass increases, these bands can mix, and the Dirac points disappear. However for some Dirac ribbons, the Dirac points move toward the Brillouin zone edge, and can fuse into one, before becoming massive, as can be seen in Fig. \ref{figsl}(c) by the green dotted line.

Finally, before constructing an effective theory of the ribbons, we first briefly explore the relationship between the gradient of the gap as a function of $m$, ie. $\partial\epsilon_{BG}/\partial m$ of Dirac ribbons, and the location of the domain wall, $i_W$. A quasi-linear scaling law is observed, but ribbons where $i_{W}=3n+2$ and ribbons where $i_{W}=3n+3$ have the same gradient (for a given $n$). We find three scaling relationships here, one where the $i_W$ is placed such that the smaller domain itself makes a Dirac ribbon, one where it makes a GS ribbon, and one where it makes a PG ribbon. Respectively, the three scaling laws are given by $\partial\epsilon_{BG}/\partial m\bigr|_{m=0}=(1-2\eta)-\alpha\eta\frac{i_W}{W}$. where, $\alpha=0,1,2$, and $\eta=1/(W+1)$

To understand the variation of gap with respect to the mass, we introduce a two-species low energy effective model for ribbons with Hamiltonian

\begin{equation}
H = H_Lc_L^\dag c_L + H_R c_R^\dag c_R + (H_{LR}c_L^\dag c_R + \mathrm{c.c.})
\label{Heff}
\end{equation}

Where $L$ and $R$ denote the left and right species, with Hamiltonians given by

\begin{equation}
H_i(k) = k\sigma_x + (s_i m + k^i_n)\sigma_y + t'_i\sigma_z,
\end{equation}

and $H_{LR} = t\sigma_0$. Here $k$ is along the ribbon axis and so is a good quantum number, $s_i m$ is the mass from equation \ref{pot} where $s_i = \pm 1$ for $i = \mathrm{L,R}$. The parameters are justified as follows. $k^i_n$ is the gap obtained from the boundary condition in the original notation of Brey and Fertig  (with the extra index $i$ denoting left or right species), which is simply a quantized $k_y$ obtained by taking a finite square well boundary condition and admixing the two sublattices. $t$ is the overlap of the wavefunctions in the two sub-systems, which is given by $t = \int_0^W dy \psi_L^\dag(r_y) H \psi_R(r_y)$ where $H$ is the original Hamiltonian from equation. $t'_i$ is an extra \emph{intra}-species overlap integral caused by the overlap of each species in the \emph{other} subsystem, such that, for example $t_L' = \int_{i_W}^W dy \psi_L^\dag(r_y) H_R \psi_L(r_y)$. This latter term is interesting both in that it represents inter-species coupling in another system, but also that it produces a mass which is orthogonal to the original mass terms $s_i m + k_n$.

The mass term $s_R m + k_n$ causes the decrease in the band gap since $s_R = -1$. This mass term changes sign when $m = k_n^L$, and beyond this value the two species have the same sign of their $\sigma_y$ mass term.

The lowest eigenvalues of equation \ref{Heff} at $k=0$ give the gap equation such that $\Delta = 2\mathrm{min}(\lambda_i)$ where $\lambda_i$ are the eigenvalues, and these can be easily solved analytically. In particular we show three specific results. Firstly, for $k_n^L = k_n^R = t'_L = t'_R =  0$, we have

\begin{equation}
\Delta(m,t) = 2(m^2 + t^2)^{1/2}
\end{equation}

Which is the monotonic increasing case found in Fig. \ref{massdependence}(c) for the regular massless armchair ribbons. Secondly, for $k_n^L = k_n^R$ finite but still $t'_L = t'_R = 0$, we obtain the similar result

\begin{equation}
\Delta(m,t,k_n) = 2\bigl|k_n - (m^2 + t^2)^{1/2}\bigr|
\end{equation}

Which corresponds to massive armchair ribbons with an equal width in each sub-system. The vanishing of the intra-species overlap $t'$ terms in this case has been confirmed numerically. Finally, we show the gap equation for $t'_L = t'_R$ and $k_n^L = k_n^R$ which is slightly unphysical, but is simpler to look at than the general case while still containing all the essential elements:

\begin{equation}
\begin{split}
\Delta(m,t,k_n,t') = &\bigl(k_n^2 + m^2 + t^2 + t'^2 \\
&- 2(k_n^2m^2 + k_n^2t^2 + t'^2t^2)^{1/2}\bigr)^{1/2}
\end{split}
\end{equation}

The effect of the non-zero $t'$ terms, which corresponds to unequal widths of the two sub-systems, is to produce a gap at $m = \sqrt{k_n^2+t^2}$ where $\Delta(m,t,k_n)$ was gapless.

So we find that the effective model reproduces all the low energy physics of the larger armchair ribbon system.

These results should prove useful in understanding the physics of several experimentally realized materials. The first is adsorbed graphene. Graphene has been successfully hydrogenated and fluorinated, which has produced some very interesting results \cite{hyd,flu}. However a convincing physical model for either has thus far proven elusive. The results outlined in this work are particularly relevant to adsorbed graphene. It has been shown that randomly adsorbed graphene will tend toward uniform adsorption. In particular in the case of semi-hydrogenated graphene, it has been shown that hydrogen atoms will tend to bond with just one of the two sublattices. However, it is extremely unlikely that randomly adsorbed graphene will become completely uniformly adsorbed, and instead should form domains, just as those proposed here. Although the domain wall configuration presented here is ideal, the variation of the mass-gap as a function of domain wall position should prove extremely useful for identifying the behaviour of different adsorbed samples. Further, via controlled substrate blocking, it is possible that the adsorption could be tunable in different samples. If possible, a two-gap system could be fabricated, even where the smaller gap could be vanishingly small.

Other systems of particular relevance to these results are twisted bilayer graphene, and substrate or lattice-mismatched graphene \cite{slg}. The idea behind these systems is that bilayer graphene, or graphene atop a substrate, behaves as a single layer of graphene with a gauge field \cite{blgmoire}. A twisted bilayer, for example, can be described as a single layer of graphene with an oscillating mass term (and gauge field which does not contribute to the gap). As the mass term changes sign, it forms a network of domain walls over the sample. It has been suggested that these domain walls \emph{always} lie along an armchair edge \cite{timotony}. Therefore this system forms precisely the systems discussed above, where the magnitude of the mass depends on the interlayer spacing. Other forms of lattice-mismatched bilayer graphene, as well as substrate mismatched graphene behave analogously.

In conclusion, we identified two geometric classes of armchair ribbon of period three in the width other than the usual Dirac class, namely the potentially gapless class, and the gapless-superlattice class. We have shown, by constructing an effeective theory of these materials, that due to the competing masses of the sublattice potential and the boundary conditions, an ordinarily massive ribbon's gap can be closed, and an ordinarily massless ribbon's gap can be opened. It is crucial for future experimental investigations, to understand that ribbons of \emph{any} width can be made gapless, and that bilayer systems in particular can be shown to have effective single layer behaviour that \emph{necessarily} falls into one of these three classes. Finally we proposed several systems which have already been realized experimentally to which these results are particularly relevant, and we believe should be very useful.

\end{document}